\pdfoutput=1
\documentclass{emulateapj}
\usepackage{xcolor}

\newcommand{\Msun}{M$_{\odot}$}
\newcommand{\peryr}{\hbox{yr$^{-1}$}}

\newcommand{\Cdoublet}{{\rm C}\kern 0.1em{\sc iii}]~$\lambda \lambda 1907$,1909}

\slugcomment{Accepted for publication in ApJ Letters}

\shorttitle{C III] emission in star-forming galaxies}
\shortauthors{Rigby et al.}

\begin{document}


\title{C III] Emission in Star-Forming Galaxies Near and Far}


\author{
J.~R.~Rigby\altaffilmark{1}, 
M.~B.~Bayliss\altaffilmark{2,3}
M.~D.~Gladders\altaffilmark{4,5}, 
K.~Sharon\altaffilmark{6}, 
E.~Wuyts\altaffilmark{7}, 
H.~Dahle\altaffilmark{8}, 
T.~Johnson\altaffilmark{6}, \&
M.~Pe\~na-Guerrero\altaffilmark{9}
}

\altaffiltext{1}{Astrophysics Science Division, 
           Goddard Space Flight Center, 8800 Greenbelt Rd., Greenbelt, MD 20771}
\altaffiltext{2}{Department of Physics, Harvard University, 
                 17 Oxford St., Cambridge, MA 02138}
\altaffiltext{3}{Harvard-Smithsonian Center for Astrophysics, 
                 60 Garden St., Cambridge, MA 02138}
\altaffiltext{4}{Department of Astronomy \& Astrophysics, University of
           Chicago, 5640 S. Ellis Ave., Chicago, IL 60637}
\altaffiltext{5}{Kavli Institute for Cosmological Physics, University of
          Chicago, 5640 South Ellis Ave., Chicago, IL 60637}
\altaffiltext{6}{Department of Astronomy, University of Michigan, 
          500 Church St., Ann Arbor, MI 48109}
\altaffiltext{7}{Max Plank Institute for Extraterrestrial Physics, 
                 Giessenbachstrasse 1, 85748 Garching, Germany}
\altaffiltext{8}{Institute of Theoretical Astrophysics, University of Oslo, 
              P.O. Box 1029, Blindern, NO-0315 Oslo, Norway}
\altaffiltext{9}{Space Telescope Science Institute, 3700 San Martin Dr., Baltimore, MD 21218}


\begin{abstract}
We measure  \Cdoublet~\AA\ emission lines in eleven 
gravitationally--lensed star-forming 
galaxies at $z\sim1.6$--3, finding much lower equivalent widths 
than previously reported for 
fainter lensed galaxies \citep{Stark:2014fa}.  While it is not yet clear what causes some 
galaxies to be strong C~III] emitters, C~III] emission is not a universal
property of distant star-forming galaxies.
We also examine  C~III] emission in  46 star-forming galaxies in the local universe,
using archival spectra from GHRS, FOS, and STIS on \textit{HST}, and \textit{IUE}.
Twenty percent of these local galaxies show strong C~III] emission, with 
 equivalent widths  $<-5$~\AA. 
Three nearby galaxies show C~III] emission equivalent widths 
as large as the most extreme emitters yet observed in the distant universe; 
all three are Wolf-Rayet galaxies.
At all redshifts, strong C~III] emission may
pick out low-metallicity galaxies experiencing intense bursts of star formation.
Such local C~III] emitters
may shed light on the conditions of star formation in certain extreme high-redshift galaxies.
\end{abstract}



\section{Introduction}
Rest-frame ultraviolet spectroscopy of star-forming $z\sim 2$--7 galaxies
has revealed strong (rest-frame equivalent width $W_r < -5$~\AA)
emission lines of \Cdoublet~\AA\
\citep{Shapley:2003gd,Erb:2010iy,Stark:2014fa,Stark:2015if}.  
\citet{Stark:2014fa} suggest that  C~III] emission may 
be a way to spectroscopically confirm the highest-redshift galaxies.  Such a tool
would be particularly important if Lyman $\alpha$ emission from galaxies 
at the reioniziation epoch is suppressed by  neutral gas absorption 
\citep{Treu:2012cc,Treu:2013ev,Pentericci:2011jn,Pentericci:2014cw,Faisst:2014fp,Tilvi:2014ei,
Caruana:2013fi,Caruana:2014ie}.
Such studies would need to understand selection effects---namely, the frequency 
of strong emission among star-forming galaxies.

There is a wide range in observed C~III] equivalent width at $z\sim2$ (see \citealt{Stark:2014fa} Figure~3). 
It is not clear what physical conditions cause galaxies to produce strong equivalent width C III] emission;
\citet{Stark:2014fa} find that the emission line ratios of strong C III] emitters at $z\sim2$
are consistent with low gas metallicity, high ionization parameter, and a hard radiation field.

In addition, the C~III] lines, together with lines of similar ionization potentials, namely 
O~III], N~III], N~IV], and Si~III], can measure atomic abundances 
(see study at $z=3.6$ by \citealt{Bayliss:2014ib}), and thus may 
be used to constrain nucleosynthetic yields in the early Universe.

This Letter has three objectives.
First, we measure  C~III] equivalent width in  11
gravitationally--lensed galaxies 
at $z\sim2$--3.  This increases the published set of such measurements by a third, and explores 
a different parameter space than \citet{Stark:2014fa}, namely, higher stellar mass and 
metallicity.  
Second, we measure C~III] equivalent width for star-forming galaxies in the local universe.  
For nearby samples it is easier to measure physical conditions, spatially resolve ionized regions, 
and test hypotheses for what produces strong C~III] emission.  
Moreover, if strong C~III] emission is a signature of primitive galaxies, then 
nearby galaxies with such emission should be investigated 
as  analogues of high redshift galaxies.
Third, we explore the dependence of C~III] equivalent width on gas-phase metallicity.

\section{Data and Methods}

\subsection{Measurements for distant galaxies}
We measure the equivalent widths of Ly $\alpha$ and the C~III] doublet 
in the rest-frame UV spectra of 
11 gravitationally-lensed galaxies at $1.6<z<3.1$.  
Spectra were obtained  using the MagE spectrograph \citep{Marshall:2008bsa}
 on the Clay Magellan II telescope, as  part of a large
study of the rest-frame ultraviolet spectral properties of bright lensed galaxies 
 (Rigby et al.\ in prep.)

Intrinsic (corrected for lensing magnification) 
stellar masses and star formation rates have been published for three of these galaxies
\citep{Wuyts:2012ej}, and are 3--7$\times 10^9$~\Msun\ and 20--100~\Msun~\peryr.
Since final lensing models are in development for the MagE sample, we cannot yet quote 
stellar masses and star formation rates for the others.
Metallicities from  [N II]/H$\alpha$ 
have been published for four of the sample galaxies:  
SGAS~090003.3$+$223408 \citep{Bian:2010bl},
the Cosmic Horseshoe \citep{Hainline:2009fg}, 
SGAS~J152745.1$+$065219 \citep{Wuyts:2012ej}, 
and  RCSGA 032727$-$132609 (hereafter RCS0327, \citealt{Whitaker:2014jy,Wuyts:2014eu}).
We measure the metallicity for a fifth galaxy, SGAS~J000451.7$-$010321,
from the [N II]/H$\alpha$ ratio in FIRE/Magellan spectra.
We use the second-order polynomial calibration of \citet{Pettini:2004bq}.

From the literature,  we take C~III] equivalent widths 
for six  galaxies from \citet{Erb:2010iy}, \citet{Christensen:2012dv}, \citet{Stark:2014fa}, and \citet{Bayliss:2014ib} 
with  measured metallicities. 
Four galaxies have metallicities derived by \citet{Stark:2014fa} using a 
Bayesian approach that relies 
on the strong emission lines used by the R$_{23}$ method;
the R$_{23}$ metallicity  measured by \citet{Christensen:2012dv} for one of those galaxies 
agrees with the  Bayesian method. 
\citet{Erb:2010iy} measured an R$_{23}$ metallicity for BX418 at $z=2.30$, as did
\citet{Bayliss:2014ib} for  SGAS~J105039.6$+$001730 at $z=3.625$.

\subsection{Measurements for nearby galaxies}
We take spectra of nearby galaxies from three spectral atlases:
\citet{Leitherer:2011cg},  Pe\~na-Guerrero et al.\ (in prep.), and \citet{Giavalisco:1996cp}.

For 17 galaxies from  \citet{Leitherer:2011cg},
we take metallicities from their tabulation, and measurements of C~III] 
equivalent width from \citet{Bayliss:2014ib}.
Four of these galaxies have spectra of multiple regions, totaling 27 measurements of C~III].
Thirteen 
of these galaxies have archival  \textit{International Ultraviolet Explorer (IUE)}
spectra that cover C~III], many of which show C~III] emission, 
even in favorable cases lines of moderate equivalent width, 
example IRAS~08339$+$6517.   
For the 18 galaxies of  Pe\~na-Guerrero et al.\ (in prep.), we measure   
C~III] equivalent widths from the \textit{HST} spectra, 
and take direct gas-phase metallicity measurements from their Table 9.
For the 11 galaxies (out of 22) from  \citet{Giavalisco:1996cp} 
that have continuum signal-to-noise ratio $\ga$ 1 per pixel,
we measure equivalent widths  
from archival low dispersion (6~\AA) \textit{IUE} spectra.

We measure Ly$\alpha$ equivalent widths for the subset
of  those galaxies 
that have archival spectra covering Ly$\alpha$, and redshifts sufficient,
given the spectral resolution of each spectrum, 
to cleanly separate in  the galaxy's Ly$\alpha$ emission 
from the much brighter  geocoronal Ly$\alpha$ emission  \citep{STIShandbook}.   
These criterea are met for: 
 Tol-1214-277 from  \citet{Leitherer:2011cg} (\textit{IUE} spectra);
two galaxies from Pe\~na-Guerrero et al.\ (in prep.),
IRAS~08339+6517 from \textit{HST} and \textit{IUE} spectra, 
and NGC~1741 from an \textit{IUE} spectrum; and the 
11 galaxies from \citet{Giavalisco:1996cp} (\textit{IUE} spectra).

These are spectra from published atlases, not from a complete 
galaxy sample.  Further, they were obtained using a range of aperture size:
$10\times20$\arcsec\ with \textit{IUE}, 
1--2\arcsec\ with FOS and GHRS, and 0.2\arcsec\ with STIS.
As such, some  observations capture bright star-forming regions within a galaxy,
whereas others capture the integrated galaxy spectrum.
These are necessary limitations of current ultraviolet spectroscopic atlases.

\subsection{Uncertainties}

When fitting equivalent widths of C~III] and Ly$\alpha$, 
our uncertainties incorporate statistical uncertainty as well 
as systematic uncertainty due to continuum estimation.
Each reported equivalent width is the mean of 
measurements assuming a linear continuum,  power law continuum, 
and (if the spectrum is locally flat)  median continuum.
We take the standard deviation as the systematic 
uncertainty due to continuum fitting; it is typically less than half the 
measurement uncertainty.

The spectra analyzed here vary widely in resolution and
signal-to-noise ratio, resulting in emission line equivalent width
measurements ranging from strong to weak detections.
For sources with strong emission lines and weak or non-detected continuum, 
the measurement is a lower limit on the equivalent width; the associated 
uncertainty may be large even though the emission line is very well detected, and reflects
the continuum uncertainty.

Following the defintion of equivalent width, 
equivalent widths have positive sign for absorption, and negative sign for emission.

\section{Results}
\subsection{C III] equivalent width}
\begin{figure}
\figurenum{1}
\includegraphics[width=4.in,angle=0]{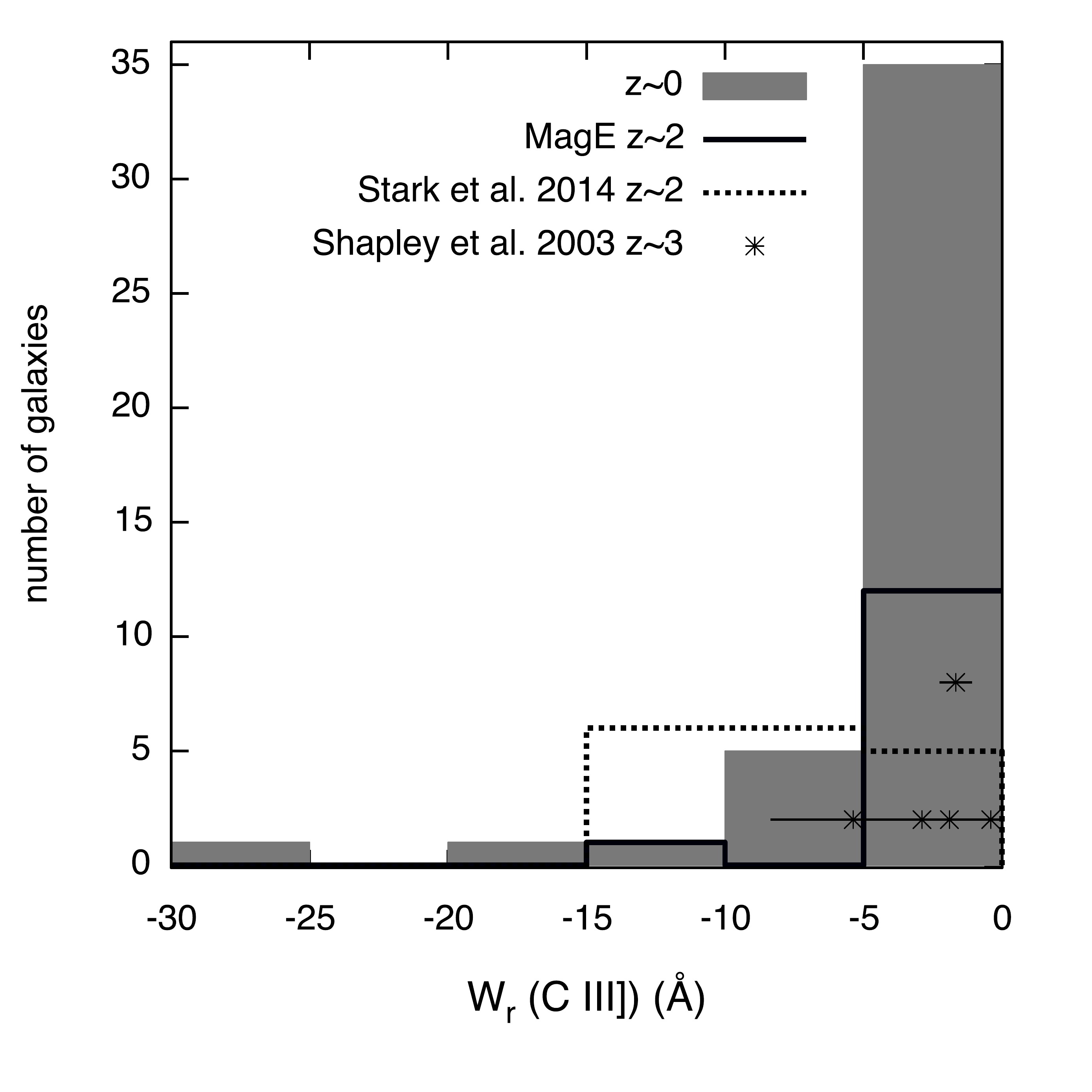}
\figcaption{Histograms of \Cdoublet~\AA\ rest-frame equivalent width.
The $z\sim0$ sample (grey filled histogram) 
combines galaxies from \citet{Leitherer:2011cg},  Pe\~na-Guerrero et al. (in prep.), and
\citet{Giavalisco:1996cp}.
Two $z\sim2$ samples are shown:  the MagE spectral atlas (solid black line), 
including the $z=3.6$ galaxy from  \citet{Bayliss:2014ib}; 
and the galaxies from \citet{Stark:2014fa} (dashed black line.)
Asterisks shows equivalent widths for the 
composite spectrum and composite quartile spectra of \citet{Shapley:2003gd},
with $1\sigma$ uncertainty, at arbitrary y values.
At both $z\sim0$ and $z\sim2$, the mode is small, and there is a tail to large equivalent widths.
}
\label{fig:histograms}
\end{figure}

We measure  C~III] strengths in bright lensed, star forming
galaxies at $z\sim2$--4, as well as in local starbursting galaxies with archival
ultraviolet spectra from \textit{HST} and \textit{IUE}.
Figure~\ref{fig:histograms} compares histograms of C~III] equivalent width 
in these samples, as well as to \citet{Stark:2014fa}  and the 
composite spectrum of \citet{Shapley:2003gd}.
Figure~\ref{fig:histograms} demonstrates that star-forming galaxies in the local universe
and at  $z\sim2$--4 show a large range of C III] equivalent width.
The mode for the distant galaxies  
and the mode for the $z\sim0$ galaxies are consistent 
with the equivalent width measured for the  \citet{Shapley:2003gd} composite.  
A significant minority of galaxies show strong C~III] emission 
with  $Wr(C~III])<-5$~\AA: 
$22\%$ of the $z\sim0$ sample,   
and $45\%$ of the $z\sim$2--4 sample.  

Indeed, the local sample includes a tail of strong C~III] emitters with
equivalent widths to  $-$25\AA, widths that are beyond that
yet observed at $z\sim2$.
These results are consistent with a scenario in which strong C~III] emission is 
not a particular feature of  $z\ga 2$ galaxies, 
but rather occurs in a minority of starbursting galaxies independent of redshift.

The MagE sample has systematically lower
equivalent widths than the \citet{Stark:2014fa} sample.
The MagE sample galaxies were selected as lensed galaxies that are most amenable to 
deep rest-frame UV spectroscopy; they have typical apparent magnitudes of $g_{AB} \sim 20$--21, 
and should be biased toward high surface brightness,
high magnification, high luminosity, and compact morphology.   
\citet{Wuyts:2012ej} examined in detail the spectral energy distributions of three galaxies
from the MagE sample, finding them typical of Lyman Break Galaxies, though younger than average.
By contrast, the \citet{Stark:2014fa} sample is selected to be much fainter,
with an average apparent V-band magnitude of $m_{AB} = 23.9$.

\subsection{C~III] equivalent width as a function of metallicity}
\begin{figure}
\figurenum{2}
\includegraphics[width=4in,angle=0]{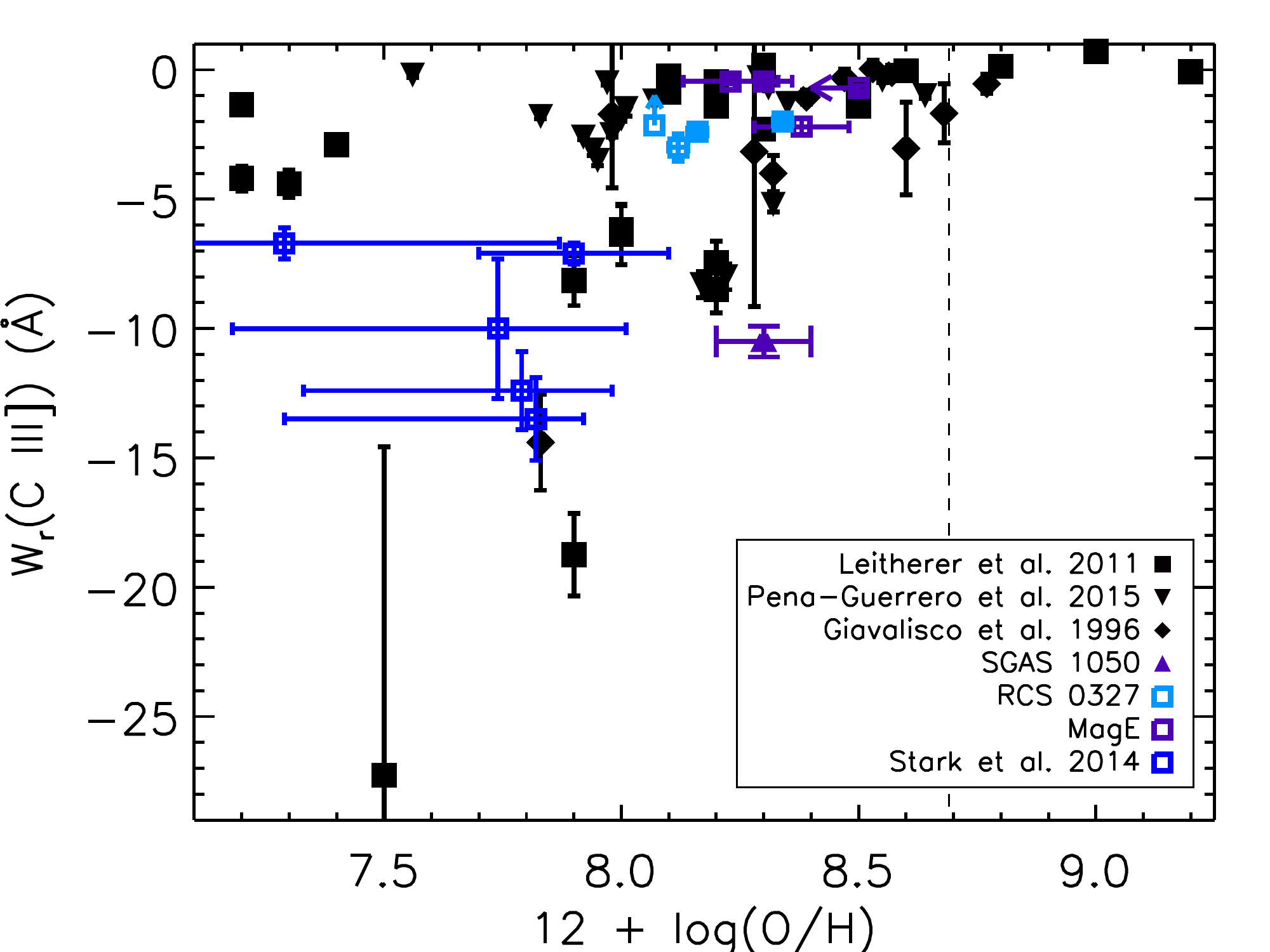}
\figcaption{Rest-frame C~III] equivalent width as a function of gas-phase metallicity.  
Nearby galaxies are plotted in black:  
\citet{Leitherer:2011cg} \textit{(black squares)};  
the STIS sample of Pe\~na-Guerrero et al. (in prep.) \textit{(black triangles)};  
and \citet{Giavalisco:1996cp} \textit{(black diamonds)}.
Redshift $z\sim2$--4 galaxies are plotted with colored symbols: 
SGAS~J105039.6$+$001730 at $z=3.6$ \citep{Bayliss:2014ib} \textit{(purple triangle)}; 
four physically-distinct star-forming regions within lensed galaxy RCS0327 at $z=1.70$ \textit{(light blue squares)};
four additional galaxies at $1.6<z<2.8$ with MagE spectra and measured metallicities \textit{(purple squares and upper limit)}; 
four galaxies from \citet{Stark:2014fa} with measured metallicity,  
and one galaxy (BX418) from \citet{Erb:2010iy} \textit{(blue squares)}.
A dashed line marks solar metallicity \citep{Asplund:2009eu}.
At large equivalent width, systematic uncertainty in the continuum level dominates.
Metallicity appears to set an envelope; low metallicity may be a necessary but 
not sufficient condition for  high C~III] equivalent width.
}
\label{fig:EW}
\end{figure}

Figure~\ref{fig:EW} plots the C~III] equivalent width as a function of gas-phase metallicity, 
for the $z\sim0$ samples (black symbols) and the 
$z\sim2$--4 samples (blue and purple symbols).  
\citet{Stark:2014fa} suggest that low metallicity may be largely 
responsible for large equivalent widths.  Figure~\ref{fig:EW} reveals that the relationship
with metallicity is complex. 
Strong emission ($\la$ -5\AA) is observed in galaxies over a wide range of metallicity, 
from 6\% to 40\% of solar.   
Strong emission appears not to occur at metallicities above $12 + \log(O/H)= 8.4$.
Indeed, only one of the six lowest-metallicity $z\sim0$ galaxies show strong emission.  
Thus, low metallicity may be a necessary but not sufficient condition for C~III] emission.

\subsection{Strong C~III] emitters in the local universe}
\begin{figure}
\figurenum{3}
\includegraphics[width=3.3in,angle=0]{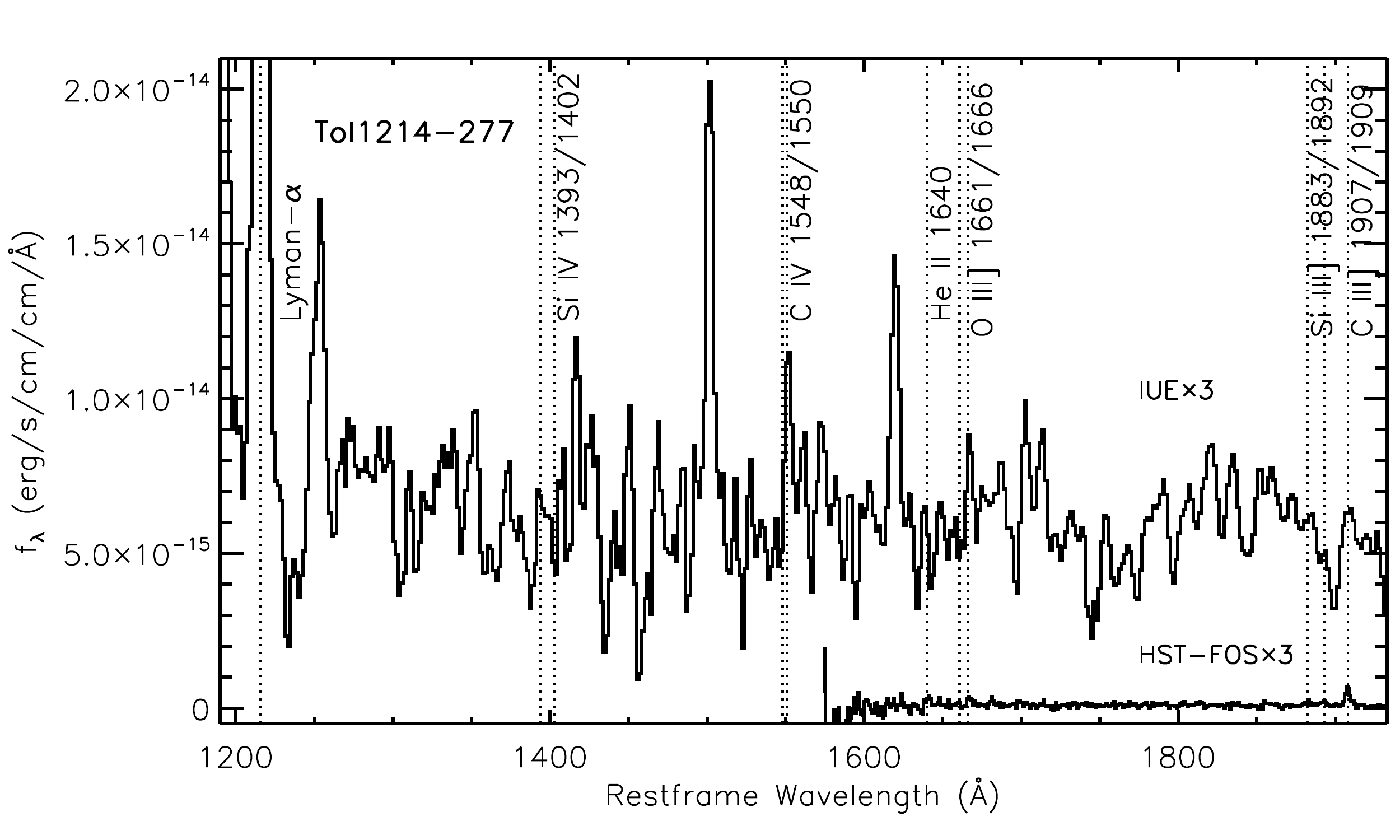} 
\includegraphics[width=3.3in,angle=0]{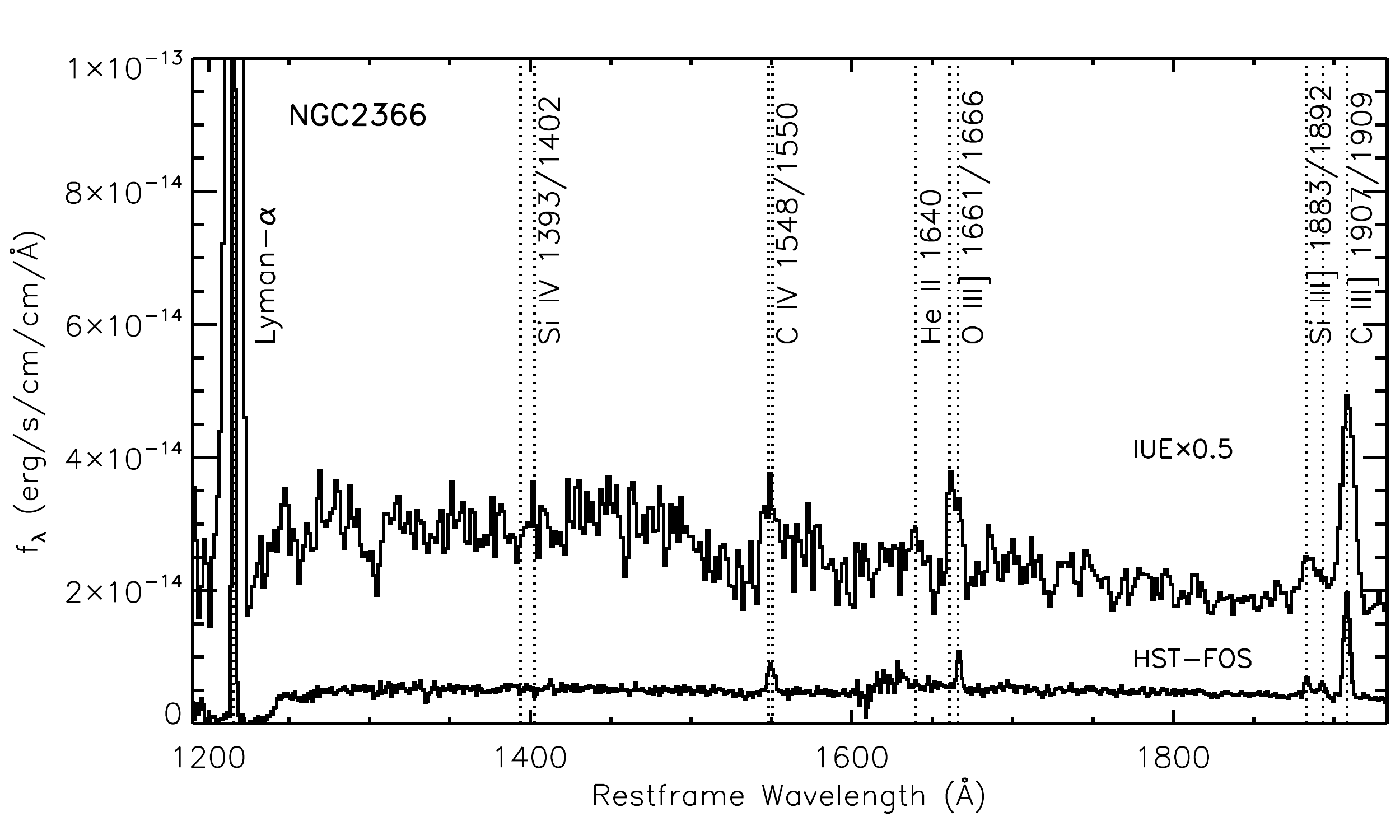} 

\vspace{0.1in}

\includegraphics[width=3.3in,angle=0]{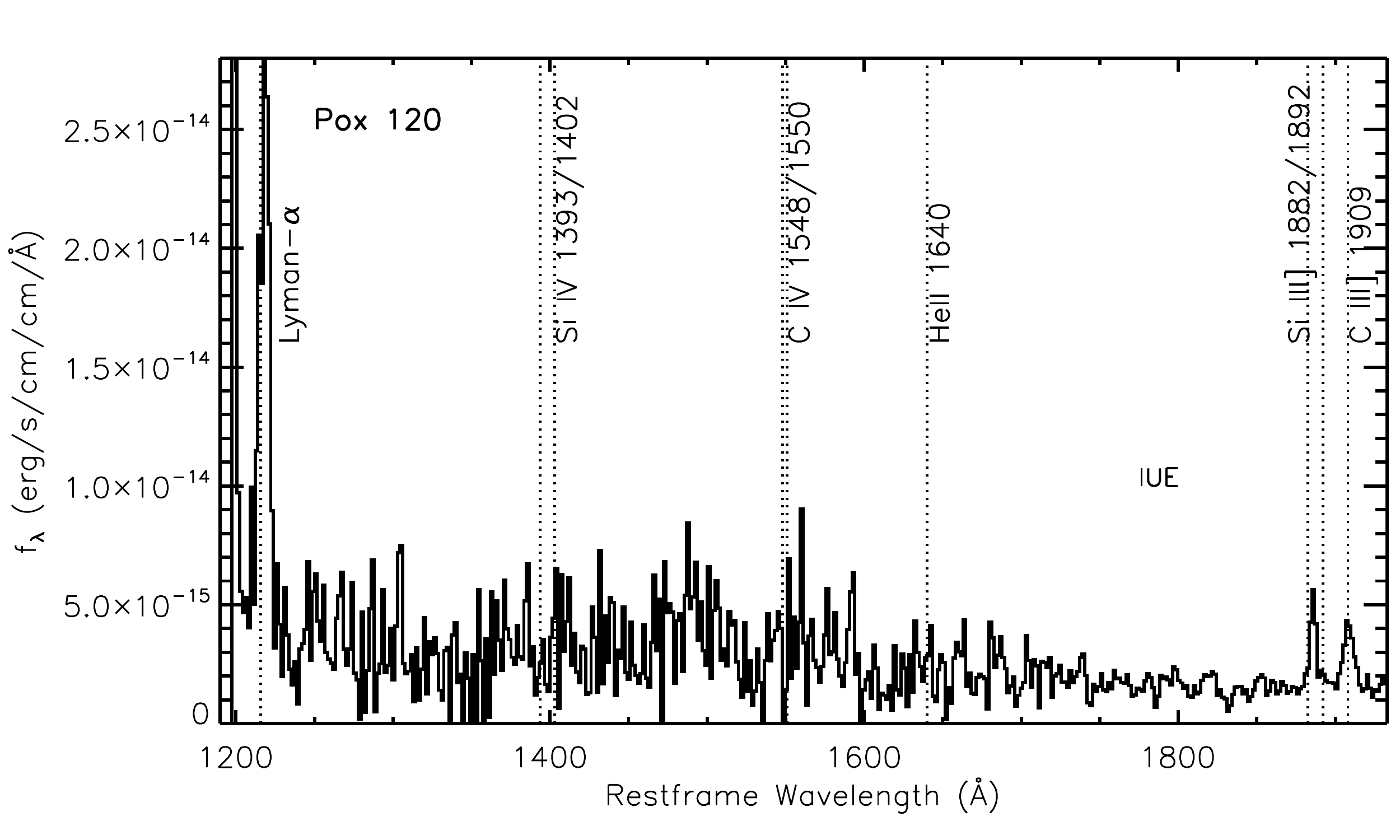}          
\includegraphics[width=3.3in,angle=0]{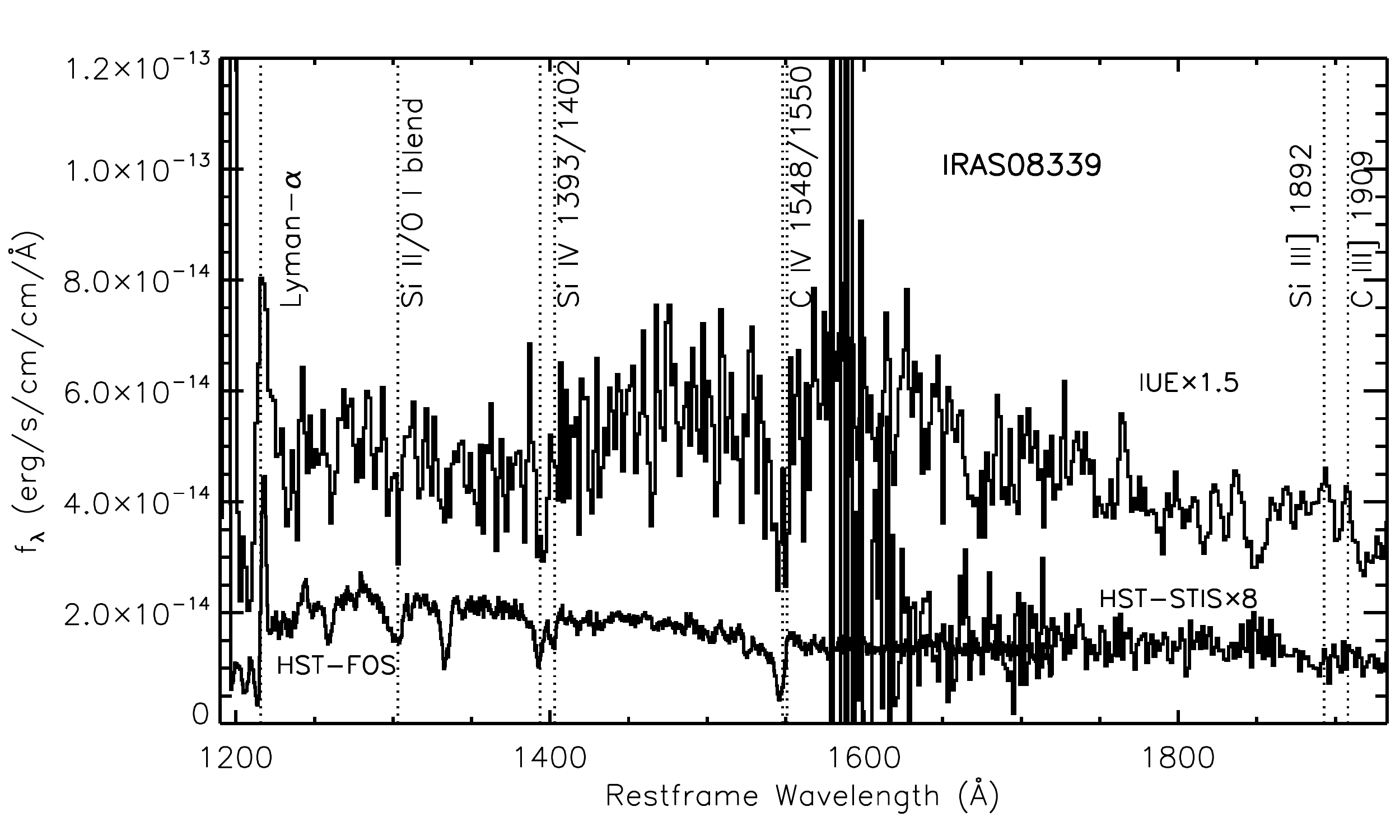}   
\figcaption{Spectra of C~III] emitters at $z\sim0$.  
The three $z=0$ emitters with largest $W_r(C~III])$  are plotted:  
Tololo $1214-277$ (top left), 
the  region Mrk 71 within galaxy NGC~2366  (top right), and Pox~120 (bottom left).  
For comparison, we also plot a spectrum with moderate emission, IRAS~08339$+$6517 (bottom right). 
The $W_r(C~III])$ for Tol~1214$-$277 is larger in \textit{HST}-FOS compared to \textit{IUE}, 
which we attribute to aperture effects.
Clearly, low-redshift galaxies can produce strong C~III] emission.
}
\label{fig:spectra}
\end{figure}

Figure~\ref{fig:EW} demonstrates that the local universe contains galaxies 
whose C III] emission is as strong, indeed stronger, than yet observed at $z\sim2$.
The three galaxies from the $z\sim0$ samples
with the most extreme C~III],  with equivalent widths exceeding $-14$\AA,
are Tololo $1214-277$, 
Markarian~71 (a star-forming region within NGC~2366),\footnote{Many studies incorrectly name this object; 
see http://cdsarc.u-strasbg.fr/viz-bin/vizExec/.getpuz?7239\&N\%202363}  and Pox~120.
We now explore what is known about these three galaxies, 
and plot their ultraviolet spectra in Figure~\ref{fig:spectra}.  

Tololo $1214-277$ is classified as a Wolf-Rayet galaxy; 
Mrk~71 is classified as a star cluster containing Wolf-Rayet stars within the irregular galaxy NGC~2366, 
and Pox~120 is classified as a Wolf-Rayet galaxy \citep{Kunth:1985vk,Schaerer:1999ku}.
The equivalent width of H$\alpha$ is extremely high for  
Mrk 71 
 ($<-1000$\AA, \citealt{Buckalew:2005ch}), 
and for TOL 1214-277 ($-1700$\AA, \citealt{Guseva:2011dj}).
We could not find a published $W_r(H\alpha)$ for Pox 120. 
Only 1 (15) of the 417  nearby galaxies of \citet{Moustakas:2006hv} shows 
an H$\alpha$ equivalent width of $<-1000$\AA\  ($<-$200\AA).
Simple stellar populations with such high equivalent widths have ages of 
only a few million years (c.f. Figure~23 of \citealt{Leitherer:2014ia}.)
Another indication of the extreme star formation occurring in these galaxies is 
the fact that Tololo $1214-277$ is one of very few galaxies in the local universe
that show indications of leaking Lyman continuum photons \citep{Verhamme:2015ce}.

In addition, 
 7 $z\sim0$ galaxies show strong C III] emission with   $-14 < W_r < -5$~\AA.
These are:  
NGC~5253 (regions HIIR-1 and HIIR-2), 
NGC~4861, 
NGC~5457 (region NGC 5471),  
Tololo~1345-420, 
 Sbs 1319+579, 
III Zw~107, and 
Arp 252.

\subsection{Lyman $\alpha$}
\begin{figure}
\figurenum{4}
\includegraphics[width=3.5in,angle=0]{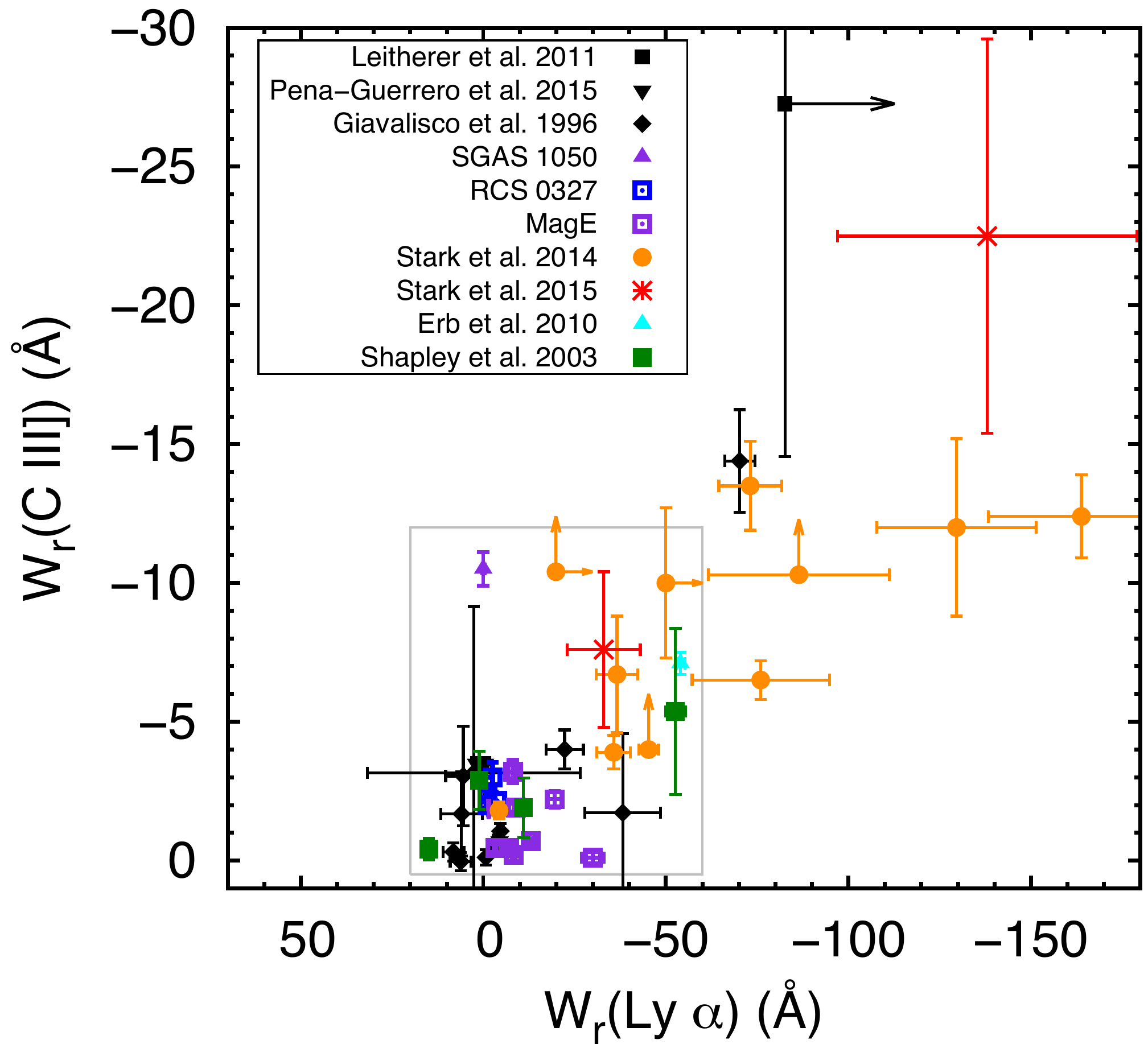}
\includegraphics[width=3.6in,angle=0]{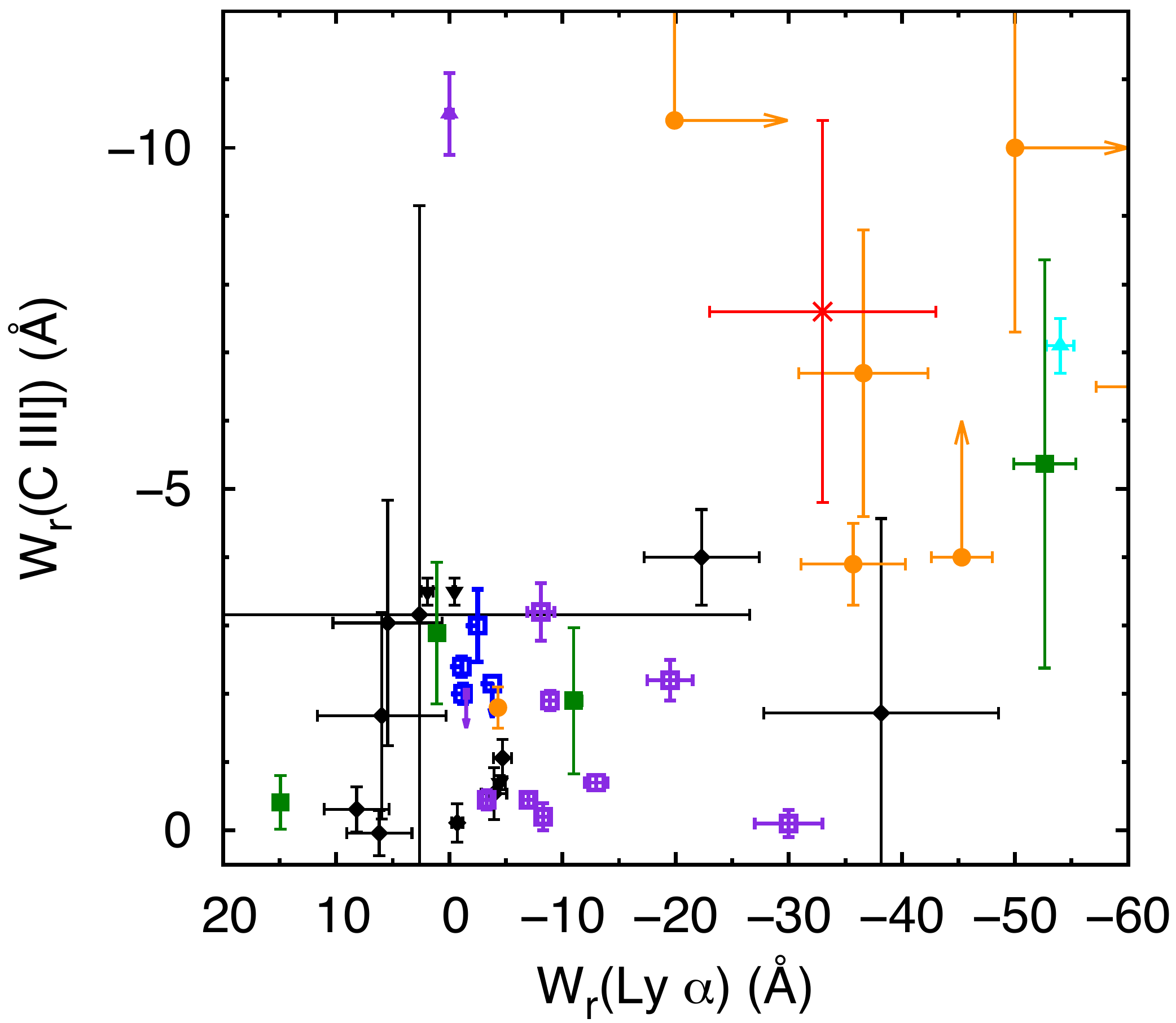}
\figcaption{ \textit{Left panel:} Comparison of rest-frame equivalent widths of Ly$\alpha$ and C III].
Nearby galaxies are plotted in black, and $1.6<z<7$  galaxies are plotted with other colors.
Symbols are coded as in Figure~\ref{fig:EW}, with additional 
values from \citet{Stark:2014fa} \textit{(orange circles)}, \citet{Stark:2015if} \textit{(red asterisks)}, 
\citet{Erb:2010iy}  \textit{(cyan triangle)}, and  \citet{Shapley:2003gd} \textit{(green squares)}.
Two  \textit{IUE} spectra were available for each of IRAS~08339+6517 and NGC~1741; 
we  plot a point for each spectrum. 
\textit{Right panel:}  Zoom of the region outlined in grey, containing most of our $z\sim2$ MagE and $z\sim0$
measurements.  
The equivalent widths are correlated, but the correlation is not obvious within the MagE sample.
}
\label{fig:Lya}
\end{figure}

\citet{Shapley:2003gd} noted a correlation between the
equivalent widths of  C III] and Lyman $\alpha$ emission in 
their quartile composite spectra.
\citet{Stark:2014fa} and \citet{Stark:2015if} further examined this correlation.
In Figure~\ref{fig:Lya}, we plot $W_r(C~III])$ versus $W_r(Ly\alpha)$ for 
galaxies at $1.6<z<7$ as well as galaxies at $z\sim0$.
A correlation is apparent for both redshift ranges, though there is considerable scatter.
The correlation is dominated by the strongest emitters,  
with W(Ly$\alpha$) $\la -50$~\AA, W(C III]) $\la -5$~\AA.
Within the dynamic range of the MagE sample, no correlation is apparent.

\subsection{Strongly star-forming galaxies that are {\bf not} strong C~III] emitters}
Since the physical conditions of the lensed galaxy RCS0327 at $z=1.70$ 
have been measured precisely, we compare them to the derived physical parameters 
of the \citet{Stark:2014fa} sample.
RCS0327 has a measured ionization parameter $\log U = -2.84 \pm 0.06$   
(\citealt{Rigby:2011il}, using the average extinction from \citealt{Whitaker:2014jy}.)
This value is consistent with ionization parameters measured for other
lensed Lyman Break galaxies:  
SGAS~J122651.3$+$215220 and SGAS~J152745.1$+$065219 in the MagE sample \citep{Wuyts:2012ej}, 
the Cosmic Horseshoe and the Clone \citep{Hainline:2009fg},
and cB58 \citep{Pettini:2002gb}.
The ionization parameters of these LBGs 
are 0.6 to 1 dex lower than those inferred by  \citet{Stark:2014fa} for four galaxies.
RCS0327 has a measured stellar mass of $6.3 \pm  0.7 \times 10^9$~\Msun, and a star formation 
rate of 30--50~\Msun yr$^{-1}$ \citep{Wuyts:2012ej}.  
Its specific star formation rate (sSFR) is $5\times 10^{-9}$~yr$^{-1}$,  
such that 
it lies a factor of 3 above the correlation between star formation rate and stellar mass at that redshift. 
Three objects in \citet{Stark:2014fa} have lower sSFRs than this; 
all have very strong C III] emission  with $W_r \le -10$\AA.
Of the four galaxies in \citet{Stark:2014fa} with highest stellar mass, $8.9<\log[M_* / M_{\odot}] <9.2$,
three have C III] equivalent widths  $<-4$\AA.
Thus, RCS0327 is similar to the  \citet{Stark:2014fa} sample in terms of sSFR, 
but has a higher metallicity and lower ionization parameter than inferred for that sample.  
This comparison provides some support for the hypothesis of \citet{Stark:2014fa}, that both 
low metallicity and high ionization parameter are required to create strong C~III] emission.

Perhaps more surprising is the fact that the 
vigorously star-forming, very low metallicity starburst galaxy 1 Zw 18 is \textit{not} 
a strong C~III] emitter; its equivalent width in three different regions varies from $-1.3$ to $-4.4$~\AA. 
Its ionization parameter is high for the local universe ($\log U \sim -2.5$ from \citet{Dufour:1988eq};
inferred as $-2.2$ as the lowest--metallicity object in \citealt{MoralesLuis:2014bj}).
Any hypothesis that explains strong C~III] emission will need to explain why 1 Zw 18 
shows relatively weak C~III] emission.

\section{Looking Forward}
C~III] emission has emerged as a practical method to spectroscopically confirm galaxies at the reionization
epoch, should Ly$\alpha$ emission prove to be suppressed. 
Our MagE spectra of lensed $1.6<z<3$ star-forming galaxies show that C~III] emission
typically has modest equivalent width in bright, high surface brightness galaxies at that epoch; 
this result is consistent with the composite spectra of \citet{Shapley:2003gd}.
Strong C~III] emission is clearly not a universal feature of galaxies at that epoch; 
equivalent widths exceeding $-5$~\AA\ are seen in $45\%$ of the $z\sim 2$--4 galaxies examined here.

We have examined C~III] emission in local star-forming galaxies with 
archival \textit{HST} spectra; 
a minority  ($22\%$)  show strong C~III] emission.   Three of these nearby galaxies show 
C~III] emission as strong or stronger ($<-14$~\AA) than yet reported for any galaxy at $z\sim2$.
The distribution function of equivalent width of C~III] emission at any redshift may contain a tail 
toward very high equivalent widths; if so, by compiling spectra for  46 nearby 
galaxies, we have sampled this tail.  The high equivalent width tail 
may be more populated at $z\sim2$ than at $z\sim0$, but determining this would require 
matched samples at both redshifts, which has not yet been done.  
The sample of \citet{Stark:2014fa} clearly samples this tail of strong emission, 
while the MagE galaxies reported here do not.

The question then becomes, what  physical conditions cause galaxies to be strong C~III] emitters?
For local galaxies, we see that metallicity may  set an envelope on maximum C~III] equivalent width.
While low metallicity galaxies show a wide range on C~III] equivalent width,
strong  emission is only seen at low metallicity.
Additional physical conditions must be 
required to produce strong C~III] emission.
The three $z\sim0$ galaxies with extreme C~III] emission, as strong as 
anything seen in the distant universe, are 
Wolf-Rayet galaxies with extremely high H$\alpha$ equivalent widths.   
Such extreme emission calls to mind the 
candidate $z\sim8$ dropout galaxies whose \textit{Spitzer}/IRAC colors are best explained by 
extreme equivalent widths of [O III] $\lambda\lambda$4959,5007   and H$\beta$ \citep{Labbe:2013kx}.
Follow-up investigations of these nearby, low-metallicity extreme C~III] emitters may thus
reveal the physical conditions that foster such strong emission from star-forming galaxies, 
with consequences for our understanding of star formation at the highest redshifts.



\acknowledgments
This paper includes data gathered with the 6.5 meter Magellan Telescopes located at 
Las Campanas Observatory, Chile.  
This research has made use of the NASA/IPAC Extragalactic Database (NED) which is operated by 
the Jet Propulsion Laboratory, California Institute of Technology, under contract with the 
National Aeronautics and Space Administration.   
This paper includes observations made with the NASA/ESA Hubble Space Telescope and with the 
International Ultraviolet Explorer, obtained 
from the Data Archive at the Space Telescope Science Institute, which is operated by the 
Association of Universities for Research in Astronomy, Inc., under NASA contract NAS 5-26555. 
JR thanks the ``First Carnegie Symposium in honor of Leonard Searle''.
We thank G.~Sonneborn and S.~Heap for  advice regarding \textit{IUE} spectra.
HD acknowledges support from the Research Council of Norway.


\begin{deluxetable}{llllllllllrll}
\tablehead{\colhead{name} & \colhead{z} & \colhead{Z} & \colhead{ref} & \colhead{$W_r(Ly~\alpha)$} & \colhead{$\sigma$} & \colhead{ref} & 
\colhead{$W_r(C~III])$} & \colhead{$\sigma$} & \colhead{ref} & \colhead{sample} & \colhead{m$_{AB}$} & \colhead{ref}\\
\colhead{} & \colhead{} & \colhead{} & \colhead{} & \colhead{(\AA)} & \colhead{(\AA)} & \colhead{} & \colhead{(\AA)} & 
\colhead{(\AA)} & \colhead{} & \colhead{}  & \colhead{} & \colhead{}
}
\startdata
RCSGA~032727$-$13260 knotE & 1.703745 & $8.34\pm0.02$ & W14 & $>-1.2$ & -99 & R14 & $-2.0$ & 0.14 & R14 & MagE & $g=19.15$ & W10\\
SGAS~J0004$-$0103 & 1.6811 & $8.30^{+0.06}_{-0.08}$ & FIRE & $-3.3$ & 0.4 & R14 & $-0.45$ & 0.14 & R14 & MagE  & $g=19.91$ & A09\\
Pox~120 & 0.020748 & 7.83 & G96 & -70.2 & 4.11 & IUE & $-14.4$ & 1.8 & IUE & G96 & & \\
IRAS~08339$+$6517 & 0.019113 & 8.31 & PG15 & $-4.45$ & 0.5 & STIS & $-0.7$ & 0.1 & STIS & PG15 & & \\
NGC~4861 & 0.002785 & 7.9 & L11 & --- & --- & --- & $-8.1$ & 1.0 & FOS & L11 & & \\
\enddata
\label{tab:tabstub}
\tablecomments{Columns: 1) source name, 2) redshift, 3) $12+\log(O/H)$ metallicity, 4) reference; 
5) Rest-frame  Ly$\alpha$ equivalent width; 6) uncertainty; 
7)  reference; 
8)  Rest-frame C~III] equivalent width. Sum of both transitions when resolved; 
9) uncertainty;
10) reference; 
11) sample;
12) magnitude for MagE sample; 
13) reference.
Redshifts for MagE galaxies are from Bordoloi et al.\ 2015 (submitted to MNRAS) for RCS0327, else
from \citet{Rigby:2014hq} and in prep.   Redshifts for $z\sim0$ galaxies are from NED.
Reference code:  A09 \citet{Abazajian:2009ef}; B07 \citet{Belokurov:2007bv}; 
B10 \citet{Bian:2010bl};   G96  \citet{Giavalisco:1996cp}; 
H09 \citet{Hainline:2009fg}; K10 \citet{Koester:2010ky}; 
L11 \citet{Leitherer:2011cg}; PG15 Pe\~na-Guerrero et al. (in prep.),
R14 \citet{Rigby:2014hq}; S07 \citet{Smail:2007bq}; 
W10 \citet{Wuyts:2010gy};  W12 \citet{Wuyts:2012gb}; W14 \citet{Wuyts:2014eu}.
Table 1 is published in its entirety in the electronic edition of ApJ Letters.  
A portion is shown here for guidance regarding its form and content.
}
\end{deluxetable}

\clearpage


\clearpage

\end{document}